\documentclass[prl,twocolumn,preprintnumbers,amsmath,amssymb,superscriptaddress]{revtex4}
\usepackage{graphicx}
\begin{document}

\title{Addendum: "On the nature of the phase transition in the itinerant helimagnet MnSi", arXiv:cond-mat/0702460v1 [cond-mat.str-el]}

\author{S. M. Stishov}
\email{sergei@hppi.troitsk.ru}\affiliation{Institute for High
Pressure Physics, Troitsk, Moscow Region, Russia}
\author{A.E. Petrova}
\affiliation{Institute for High Pressure Physics, Troitsk, Moscow
Region, Russia}
\author{S. Khasanov}
\affiliation{Institute of Solid State Physics, Chernogolovka,
Moscow Region, Russia}
\author{G. Kh. Panova}
\affiliation{Russian Research Center Kurchatov Institute, Moscow,
Russia}
\author{A.A.Shikov}
\affiliation{Russian Research Center Kurchatov Institute, Moscow,
Russia}
\author{J. C. Lashley}
\affiliation{Los Alamos National Laboratory, Los Alamos, 87545 NM,
USA}
\author{D. Wu}
\affiliation{Ames Laboratory, Iowa State University, Ames, IA
50011, USA}
\author{T. A. Lograsso}
\affiliation{Ames Laboratory, Iowa State University, Ames, IA
50011, USA}

\date{\today}

\begin{abstract}
New high resolution data for heat capacity, heat capacity under
applied magnetic fields and resistivity of high quality single
crystal of MnSi are reported. Striking mirror symmetry between
temperature derivative of resistivity and thermal expansion
coefficient of MnSi is displayed. Close similarity between
variation of the heat capacity and the temperature derivative of
resistivity through the phase transition is observed. It is shown
that the heat capacity and thermal expansion coefficient of the
helical phase are not influenced by moderate magnetic field.
\end{abstract}

\maketitle

As a continuation of our effort in studying the magnetic phase
transition in MnSi we carried out new measurements of heat
capacity and electrical resistivity of high quality single crystal
of MnSi with increased resolution and accuracy. The samples were
cut by the spark erosion from the same piece of MnSi single
crystal used earlier \cite{1}. Improved results were obtained due
to increased mass of the sample, smaller temperature steps (heat
capacity), and the optimized sample dimensions and the better
temperature control (resistivity).

\begin{figure}[htb]
\includegraphics[width=86mm]{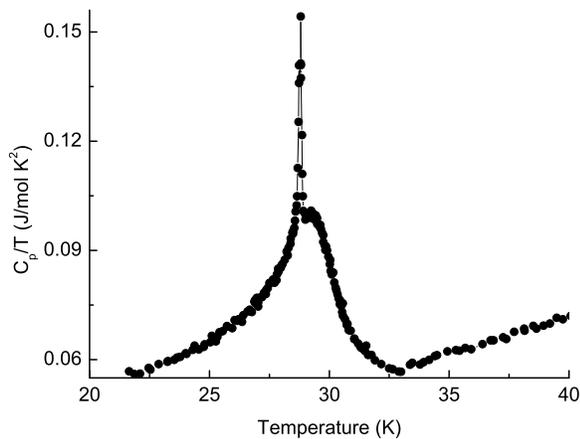}
\caption{\label{fig1} Temperature dependence of heat capacity
divided by temperature near the phase transition in MnSi }
\end{figure}
\begin{figure}[!h]
\includegraphics[width=86mm]{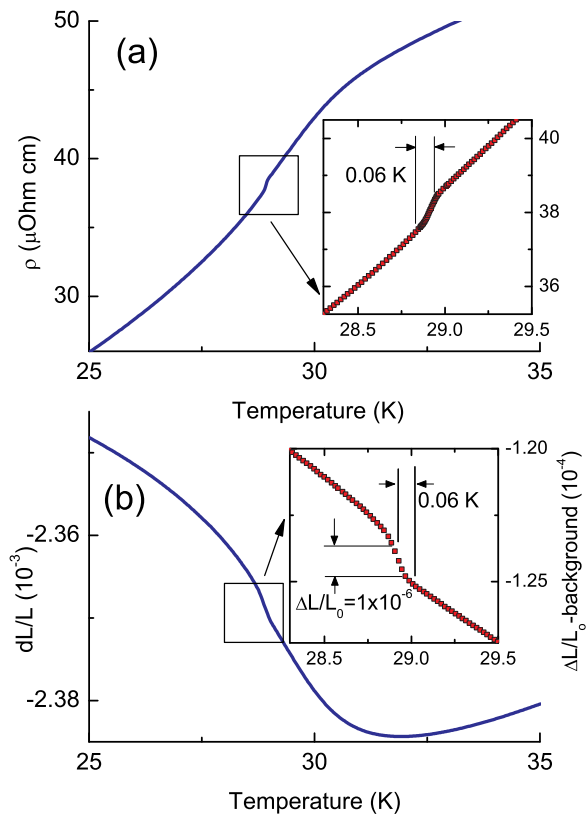}
\caption{\label{fig2}  Variations of resistivity (a) and the
relative length (b) of the sample of MnSi with temperature. Quasi
discontinuities at the transition point are shown in the insets.
For better view a background contribution was subtracted from the
original data in the inset of Fig. \ref{fig2} b. The width of the
transition is compatible with the temperature resolution of the
experiment. The linear thermal expansion is calculated by
integrating the thermal expansion coefficient (Fig. \ref{fig3})}
\end{figure}

\begin{figure}[!h]
\includegraphics[width=86mm]{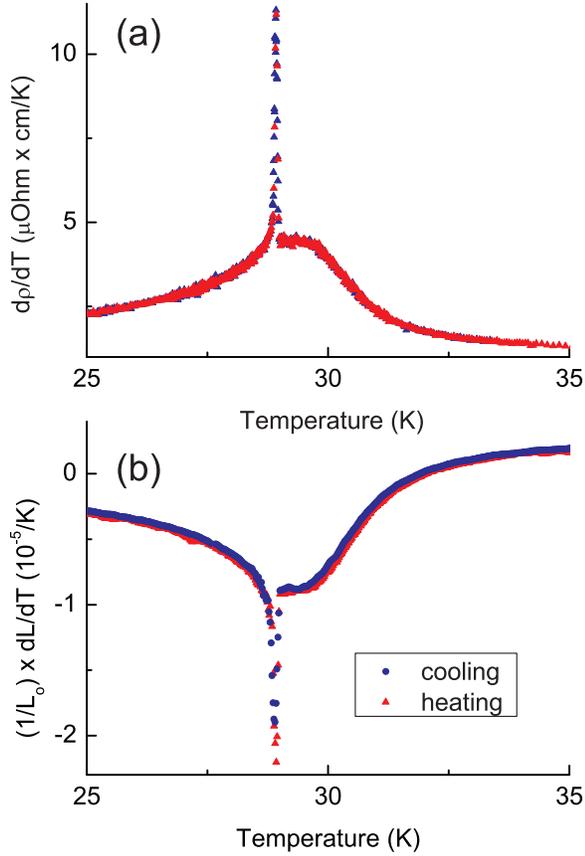}
\caption{\label{fig3} Dependence of temperature coefficient of
resistivity (a) and linear thermal expansion coefficient (b) of
MnSi on temperature in the vicinity of the phase transition }
\end{figure}
\begin{figure}[!h]
\includegraphics[width=86mm]{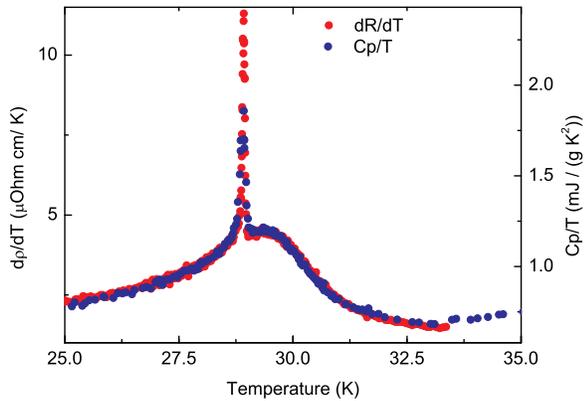}
\caption{\label{fig4} Similarity between the temperature
derivatives of resistivity and the heat capacity curves in the
phase transition region. The curves were reduced by the linear
transformation }
\end{figure}
\begin{figure}[!h]
\includegraphics[width=86mm]{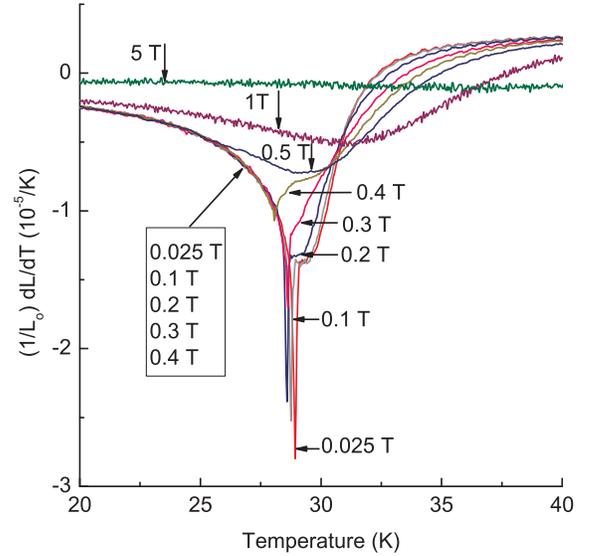}
\caption{\label{fig5} Linear thermal expansion coefficient of MnSi
near the phase transition in magnetic fields ($H\parallel[110]$).
It is seen that moderate magnetic fields up to 0.4T do not
influence thermal expansion of the helical phase}
\end{figure}
\begin{figure}[!h]
\includegraphics[width=86mm]{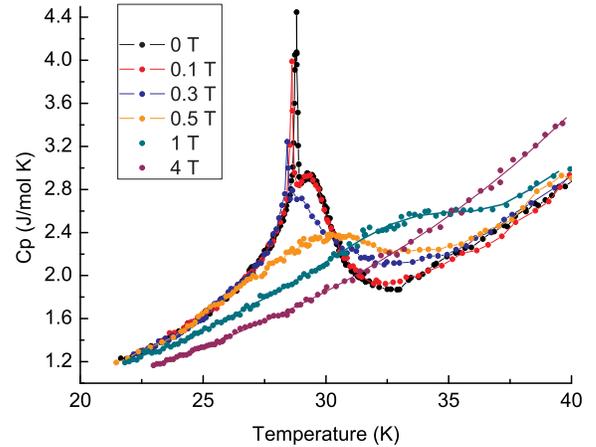}
\caption{\label{fig6} Heat capacity of MnSi in magnetic fields
($H\parallel[100]$). Note that moderate magnetic fields do not
affect much heat capacity of the helical phase.}
\end{figure}

New data on heat capacity and resistivity of MnSi are displayed in
Fig. \ref{fig1}, \ref{fig2} a, \ref{fig3} a. Like before the heat
capacity peak at the phase transition reveals a complicated
structure with sharp and broad component but the sharp component
appears to be much more prominent than it was observed earlier.
The resistivity curve (Fig. \ref{fig2} a) clearly shows a small
quasi discontinuity at about 28.9 K. The corresponding temperature
derivative of resistivity $d\rho/dT$ (Fig. \ref{fig3} a) can be
interpreted as a slightly broadened delta function developing
within the continuous anomaly, like it was pointed out before in
respect to the thermal expansion coefficient \cite{1}. Surprising
mirror symmetry between the temperature derivatives of resistivity
and the thermal expansion coefficient is illustrated in Fig.
\ref{fig3}. This obviously indicates spin fluctuations as a
dominant factor defining thermodynamic and transport properties of
MnSi in the vicinity of the phase transition. This statement is
further supported by  almost perfect similarity between the
temperature derivatives of resistivity and the heat capacity
curves in the phase transition region (Fig. \ref{fig4}) ( in this
connection see \cite{2,3,4}).

A new aspect of influence of magnetic fields at the thermal
expansion coefficient is shown in Fig. \ref{fig5}. As is seen in
Fig. \ref{fig5}, moderate magnetic fields, though do not change
nature the phase transition, strongly influence thermal expansion
of the paramagnetic phase. At the same time thermal expansion of
the helical phase does not experience even a slight change at
least up to 0.4 T. The same situation can be also seen in the heat
capacity response to moderate magnetic fields (Fig. \ref{fig6}).
This implies a significant stiffness of the helical spin structure
or in the other words a lack of extensive paramagnetic
fluctuations in MnSi even at the transition point. This situation
most probably signifies a finite value of the order parameter at
the phase transition point, therefore indicating first order phase
transition.

%


\begin{thebibliography}{99}

\bibitem{1} S. M. Stishov, A.E. Petrova, S. Khasanov, G. Kh. Panova, A.A.Shikov, J. C. Lashley, D. Wu, and T. A. Lograsso , arXiv:cond-mat/0702460v1 [cond-mat.str-el]
\bibitem{2} V.M. Nabutovskii, A.Z. Patashinskii, Fizika Tverdogo Tela, \textbf{10}, 3121 (1968)
\bibitem{3} M.E. Fisher and J.S. Langer, Phys.Rev.Lett. 20, 665 (1968)
\bibitem{4} T.G. Richard and D.J.W. Geldart, Phys.Rev.Lett. 30, 290 (1973)


\end{thebibliography}

\end{document}